\documentclass[11pt]{article}
\usepackage[dvipdfmx]{color}
\usepackage{graphicx}
\usepackage{xcolor}
  \usepackage{amsthm,amsfonts}
  \usepackage{amsmath}
\usepackage{mathabx}
\usepackage{slashed}
\usepackage{ulem}
\usepackage{hyperref}
\usepackage{cleveref}
\usepackage{cite}

\newcommand{\bea}   {\begin{eqnarray}}
\newcommand{\eea}   {\end{eqnarray}}

\def\zzg{${\mathbb Z}_2\times{\mathbb Z}_2$-graded }

\begin{document}
\renewcommand{\thefootnote}{\fnsymbol{footnote}}

\thispagestyle{empty}

\title{Braided Majorana qubits as a minimal setting\\ for Topological Quantum Computation?}
\author{ Francesco Toppan\thanks{{E-mail: {\it toppan@cbpf.br}}}
\\
\\
}
\maketitle

{\centerline{
{\it CBPF, Rua Dr. Xavier Sigaud 150, Urca,}}\centerline{\it{
cep 22290-180, Rio de Janeiro (RJ), Brazil.}}
~\\
\maketitle

\begin{abstract}
I point out that a possible minimal setting to realize Kitaev's proposal of a Topological Quantum Computation which offers topological protection from decoherence could in principle be realized by braided Majorana qubits.
Majorana qubits and their braiding were introduced in Nucl. Phys. B  980, 115834 (2022)  and further analyzed in J. Phys. A: Math. Theor. 57, 435203 (2024). Braided Majorana qubits implement a Gentile-type parastatistics with at most $s-1$ excited states accommodated in a multiparticle sector (the integer $s=2,3,4,\ldots$ labels quantum group reps at roots of unity).
It is argued that braided Majorana qubits could play, for topological quantum computers, the same role as standard bits for ordinary computers and as qubits for ``ordinary" quantum computers.
\end{abstract}
~\\
~\\
~\\
~\\
~\\
~\\
~\\
~\\
~\\
~\\
~\\
~\\
~\\
~\\
~\\

\hfill{CBPF-NF-002/25}
\newpage

\section{Introduction}

One of the fascinating areas of research is based on the Kitaev's proposal \cite{kit} (see also \cite{{nssfds},{brki}}) to use emergent Majorana particles to encode the properties of a topological quantum computation which offers topological protection under quantum decoherence. The logic of topological quantum computation with Majorana fermions has been investigated in \cite{kau} and theoretical models have been proposed, see e.g. the ``Fibonacci particles" presented in \cite{kalo}. The recent Microsoft's announcement \cite{microsoft} of the first quantum chip powered by a topological architecture points towards a practical implementation of the Kitaev's proposal (see the \cite{roadmap} roadmap to fault tolerant quantum computation and \cite{{pikulin},{inasal}} for the production of devices admitting a topological phase with Majorana zero-modes).\par
Anyonic braid statistics can only apply to emergent particles living in $2$ space dimensions. \\ In order to fulfill Kitaev's program a physical model should satisfy the following three conditions:\\
~
\\
{\it i}) at the theoretical level it should be able to accommodate a braid statistics,\\
{\it ii}) also at theoretical level, it should be able to encode the logic operations of topological quantum computation and, finally,\\
{\it iii}) it should be reproduced on a physical media/device, just like ordinary chips (or, in the old days, transistors) encode the Boolean logic operations.\par
~
\par
A relevant question which could be addressed is the following: which is the minimal theoretical setting for realizing
the program of a Topological Quantum Computation (TQC for short)? \\
To give perspective, it  makes sense to address this question within a broader picture, putting it in comparison with 
standard computers and ``ordinary" quantum computers (here, ``ordinary" stands for not topologically protected). We know that\\
~
\\
$\bullet$ standard computers manipulate bits of information via Boolean logic gates, while\\
$\bullet$ what can be called  the ``ordinary quantum computers" operate at the level of qubits, considered as the minimal building blocks. Non-minimal settings based, instead of $2$-state qubits, on qutris and more general qudits
(see, e.g., \cite{gnss}) are also possible; they received, nevertheless, much less attention for the obvious reason that
it is much easier to perform the basic quantum operations on ``minimal qubits". 
\\~\par
The above question can therefore be rephrased as: which is the analogous, for Topological Quantum Computation, of ordinary bits for Boolean logic and of qubits for operations performed by quantum computers?\par
As a possible candidate, the minimal building blocks could be realized by (braided) Majorana qubits which satisfy the above condition ${\it i}$). Majorana qubits and their braiding  (based on the \cite{kasa}  $R$-matrix of the Alexander-Conway polynomial in the linear crystal rep on exterior algebra) were introduced in 
\cite{topqubits} (for a short presentation, one can see \cite{topscipost}). Braided Majorana qubits implement a Gentile-type parastatistics \cite{gen} with at most $s-1$ excited states accommodated in a multiparticle sector (for an integer $s=2,3,4,\ldots$ which labels quantum group reps at roots of unity, with the $s=2$ case corresponding to ordinary fermions transforming under the permutation group). A significant feature of braided Majorana qubits is
that they are described \cite{topvoli} by new mathematical structures (like a generalization of {\it Volichenko algebras}, more on that below) which have yet to be fully investigated.\par
Braided Majorana qubits satisfy, for the braid statistics, the minimality criterium which applies to bits and ordinary qubits. They are therefore a natural playground  to put to test ideas and different mathematical frameworks to be applied to the braid statistics of quantum models. Braided Majorana qubits pass the $i$) condition.  One of the purposes of this paper is to draw the  attention to this novel minimal physical model  in order to start addressing its properties concerning the conditions ${\it ii}$) and ${\it iii}$) mentioned before.\\
~\par
This paper presents in a concise and unified framework the three mathematical structures that are known to describe the braided Majorana qubits.  It is based on \cite{topqubits} and the new results in \cite{topvoli}. The three mathematical structures under consideration are:\\
$\bullet$ the braiding induced by a graded Hopf algebra endowed with a braided tensor product, \\
$\bullet$ the (superselected) reps of the quantum superalgebra ${\cal U}_q({\mathfrak{osp}}(1|2))$ truncated at roots of unity,\\
$\bullet$ the generalization of {\it Volichenko algebras} inducing ``mixed-bracket" (i.e., interpolating ordinary commutators/anticommutators) Heisenberg-Lie algebras.\par

In order to focus on the main ideas, this paper illustrates the main results by skipping the demonstrations 
presented in the references cited in the text. The paper is structured as follows: Section {\bf 2} introduces the notion of a ${\mathbb Z}_2$-graded Majorana qubit, while the three mathematical structures defining the braiding are respectively presented in Sections {\bf 3}, {\bf 4} and {\bf 5}. 
Further comments (about the new mathematical structures describing braided Majorana qubits, the minimal setting and the possible application to the logic of topological quantum computation) will be given in the Conclusions. 

\section{${\mathbb Z}_2$-graded Majorana qubits}

A ${\mathbb Z}_2$-graded Majorana qubit corresponds \cite{topqubits} to a bosonic  vacuum state $|0\rangle$ and a fermionic excited state $ |1\rangle$:
{\small{\bea\label{qubit01}
|0\rangle = \left(\begin{array}{c} 1\\0\end{array}\right) , &\quad& |1\rangle =\left(\begin{array}{c} 0\\1\end{array}\right) .
\eea}}
The following operators, acting on the ${\mathbb Z}_2$-graded qubit, close the Lie superalgebra ${\mathfrak{gl}}(1|1)$ \cite{kac}:
{\small{\bea\label{4op}
&\alpha =\left(\begin{array}{cc} 1&0\\0&0\end{array}\right),\quad ~\beta =\left(\begin{array}{cc} 0&1\\0&0\end{array}\right),\quad ~ \gamma =\left(\begin{array}{cc} 0&0\\1&0\end{array}\right),\quad ~\delta =\left(\begin{array}{cc} 0&0\\0&1\end{array}\right).&
\eea}}
The defining brackets, given by (anti)commutators, are
\bea\label{anticomm}
&\relax [\alpha,\beta]=\beta, \qquad [\alpha,\gamma]=-\gamma,\qquad  [\alpha,\delta]=0,\qquad [\delta,\beta]=-\beta,\qquad  [\delta,\gamma]=\gamma,&\nonumber\\
&~\quad\{\beta,\beta\}=\{\gamma,\gamma\}=0,~\quad\qquad\qquad \{\beta,\gamma\}=\alpha+\delta.&
\eea
The diagonal operators $\alpha,\delta$ are even, while $\beta,\gamma$ are odd ($\gamma$ being the fermionic
creation operator).\par
The ${\mathbb Z}_2$-grading is given by
\bea
&{\mathfrak{gl}}(1|1)={\mathfrak{gl}}(1|1)_{[0]}\oplus{\mathfrak{gl}}(1|1)_{[1]}, \qquad {\textrm{with}} \quad \alpha,\delta\in{\mathfrak{gl}}(1|1)_{[0]}\quad {\textrm{and}}\quad \beta,\gamma\in {\mathfrak{gl}}(1|1)_{[1]}.&
\eea
The admissible nonvanishing entries of the even/odd ${\mathfrak{gl}}(1|1)$ generators are expressed by the 
``$\ast$" symbol entering the respective $2\times 2$ matrices:
{\footnotesize{
\bea
{\mathfrak{gl}}(1|1)_{[0]}\equiv \left(\begin{array}{cc} \ast&0\\0&\ast\end{array}\right), &&{\mathfrak{gl}}(1|1)_{[1]}\equiv \left(\begin{array}{cc} 0&\ast\\\ast&0\end{array}\right).
\eea
}}

The excited state $ |1\rangle$ is a Majorana fermion since it coincides with its own antiparticle. \par
The ${\mathbb Z}_2$ grading makes (\ref{qubit01}) to differ from an ordinary qubit. In particular,
bosons/fermions satisfy a superselection rule which implies that they cannot be superposed (linearly combined).\par
In the case of an ordinary qubit the inequivalent configurations, determined by the ray vectors, are expressed by the
Bloch sphere ${\mathbf S}^2$. In the case of the Majorana qubit its analogous Bloch sphere, determined by the ray vectors, is expressed by  ${\bf Z}_2$, namely just one bit of information. The identification goes as follows: $0\equiv |0\rangle$, $1\equiv |1\rangle$.

\section{First scenario of braided Majorana qubits:\\ ~~~~~~ Multiparticle First Quantization (graded Hopf algebra with braided tensors).}

The first scenario of braiding the ${\mathbb Z}_2$-graded Majorana qubits consists in introducing \cite{topqubits} a multiparticle First Quantization defined by the graded Hopf algebra ${\cal U}({\mathfrak{gl}}(1|1))$ (the Universal Enveloping Algebra of $\mathfrak{gl}(1|1)$) endowed with, following the \cite{maj} prescription, a braided tensor product $\otimes_{br}$ which is compatible with the  ${\mathfrak{gl}}(1|1)$ superalgebra.\par
In terms of the $4\times 4$ matrix $B_t$, parametrized by a nonvanishing complex parameter $t$,
the creation operator $\gamma$ defined in (\ref{4op})  is assumed to satisfy 
\bea \label{braidingamma}
({\mathbb I}_2\otimes_{br} \gamma)\cdot (\gamma\otimes_{br} {\mathbb I}_2) &=& B_t\cdot (\gamma\otimes_{br} {\mathbb I}_2)\cdot ({\mathbb I}_2\otimes_{br} \gamma) \equiv B_t\cdot (\gamma\otimes_{br}\gamma),
\eea
where the ``$\cdot$" symbol denotes the standard matrix multiplication and ``${\mathbb I}_n$" denotes the $n\times n $ identity matrix.
The matrix $B_t$ (the $R$-matrix of the Alexander-Conway polynomial in the linear crystal rep on exterior algebra
\cite{kasa}) is given by
{\footnotesize{\bea\label{btmatrix}
B_t&=&\left(\begin{array}{cccc} 1&0&0&0\\0&1-t&t&0\\0&1&0&0\\0&0&0&-t\end{array}\right).
\eea
}}
The $t\neq 0$ condition ensures that $B_t$ is invertible. \par

The ${\otimes}_{br}$ braided tensor product  defined in (\ref{braidingamma}) satisfies the required \cite{maj} compatibility  condition since $B_t$ obeys the braid relation
\bea\label{braidedrel}
(B_t\otimes {\mathbb I}_2)\cdot ({\mathbb I}_2\otimes B_t)\cdot 
(B_t\otimes {\mathbb I}_2) &=& ({\mathbb I}_2\otimes B_t) \cdot
(B_t\otimes {\mathbb I}_2)\cdot ({\mathbb I}_2\otimes B_t).
\eea

\subsection{Construction of the multiparticle sectors}

Following \cite{topqubits} the $N$-particle Hilbert space ${\cal H}_N$ is a subset of $N$ tensor products of the single-particle Hilbert space ${\cal H}$ spanned by $ |0\rangle, |1\rangle$ entering (\ref{qubit01}):
 \bea\label{subspace}
{\cal H}_{N}&\subset &{\cal H}^{\otimes N}.
\eea
${\cal H}_{N}$ is constructed by repeatedly applying, on the $N$-particle vacuum state $|0\rangle_{N}$ given by 
\bea
\qquad |0\rangle_N &=& |0\rangle \otimes \ldots \otimes |0\rangle\qquad \qquad (\textrm{$N$ times}),
\eea
the Hopf algebra coproducts of the creation operator
$\gamma$, so that ${\cal H}_{N}$ is spanned by
the normalized vectors $|n\rangle_{t,N}$, where the integer $n$ labels the $n$-th excited state:
\bea\label{spanning}
|n\rangle_{t,N} &\propto& {\widehat{\left({{ {\Delta^{(N-1)}}(\gamma)}}\right)^{n}}} |0\rangle_N, \qquad {\textrm{for $n=0,1,2,\ldots$.}}
\eea
Some comments are in order. The hat in the r.h.s. denotes the evaluation of the coproduct on the given representation of the Universal Enveloping Algebra ${\cal U}= {\cal U}({\mathfrak{gl}}(1|1))$.\\
The Hopf algebra coproduct  map $\Delta$ satisfies
\bea\label{coproductbr}
\Delta~:~{\cal U}\rightarrow {\cal U} \otimes_{br}  {\cal U},&& \Delta^{(N+1)} := (\Delta\otimes_{br} id)\Delta^{(N)}=(id\otimes_{br} \Delta)\Delta^{(N)}\in {\cal U}^{\otimes_{br} N}.
\eea
The  property
\bea\label{uaub}
   \Delta(U_AU_B)&=&\Delta(U_A)\Delta(U_B) \qquad{\textrm{for any ~$U_A,U_B\in {\cal U}$}}
\eea
implies that the action on any given $U\in {\cal U}({\mathfrak gl}(1|1))$ is recovered from the action of the coproduct on the
Hopf algebra unit ${\bf 1}$ and on the primitive elements $\zeta\in {\mathfrak{gl}(1|1)}$, respectively given by
\bea\label{deltaidg}
   \Delta({\bf 1})={\bf 1}\otimes_{br}{\bf 1}, &\quad & 
   \Delta({ \zeta})={\bf 1}\otimes_{br}{\zeta}+\zeta\otimes_{br} {\bf 1}.
\eea
The $N$-particle Hamiltonian $H_N$ can be expressed in terms of the single-particle Hamiltonian $H_1$ defined  
as $H_1:= \delta=diag(0,1)$; we have
\bea\label{nparticlehamiltonians}
H_N &:=& {{ {\widehat{ \Delta^{(N-1)}}}(H_1)}}.
\eea
With these positions we have all the ingredients to compute the braided multiparticle spectra for any $t\neq 0$.

\subsection{The multiparticle spectra}

We just limit here to present the results, referring to \cite{topqubits} and \cite{topvoli} for their derivation.

Truncations of the energy spectra of the multiparticle braided Majorana qubits are recovered for special root-of-unity values of $t$.  For all other $t\neq 0$ cases the energy spectra are untruncated. \par
In a convenient parametrization for $|t|=1$, the root of unity truncations are recovered from the position
\bea\label{gparametrization}
&t=-e^{2i\pi g}, \qquad {\textrm{with ~~$g=\frac{r}{s}$ ~~and $r,s$ mutually prime integers.}}&
\eea
The physics does not depend on the specific value of $t$, but only on the ``root of unity level" specified by the integer $s=2,3,4, \ldots$.  In the limit $s\rightarrow\infty$ one obtains $t=-1$ which produces an untruncated spectrum (a generic $t\neq 0$ which does not coincide with a root of unity produces the same untruncated spectrum).
Depending on $s$, the following classes of $N$-particle energy spectra are recovered
(the energy eigenvalues are given by integer numbers and are not degenerate):
~\par{{
{{{\bf - truncated $L_s$ level}, the $N$-particle energy eigenvalues $E$ are
\bea\label{truncatedenergy}
E &=& 0,1,\ldots, N\qquad \quad~{\textrm{for}}\quad N<s,\nonumber\\
E &=& 0,1,\ldots, s-1 \qquad {\textrm{for}} \quad N\geq s;
\eea
in this case a plateau is reached at the maximal energy level $s-1$; this is the maximal number of braided Majorana fermions that can be accommodated in a multi-particle Hilbert space;}}\par
~\par
{{{\bf - untruncated ($t=-1) ~~L_\infty$ level}, the $N$-particle energy eigenvalues $E$ are
\bea\label{genericenergy}
E &=& 0,1,\ldots, N\qquad {\textrm{for any given}}\quad N;
\eea
in this case there is no plateau and the maximal energy eigenvalues grow linearly with $N$.}}
}}
\par
As mentioned before, the level-$s$ root of unity implements a Gentile-type \cite{gen} parastatistics.
The special point $t=1$, which is the level-$2$ root of unity, gives the  ordinary total antisymmetrization of the fermionic wavefunctions and encodes the Pauli exclusion principle of ordinary fermions.

\section{Second scenario of braided Majorana qubits:\\ ~~~~~~ Roots of unity truncations from superselected quantum group reps.}

The truncations of the multi-particle spectra at roots of unity are reminiscent of the well-known special features of the quantum groups
representations at roots of unity, see \cite{lus} and \cite{dck}. On the other hand the approach of 
\cite{topqubits}
does not directly use quantum group data since the compatible braided tensor product is applied to the $\mathfrak{gl}(1|1)$ superalgebra, not its quantum counterpart. \par
The open question of deriving the spectrum of multiparticle braided Majorana qubits from quantum group data was solved in \cite{topvoli}; we briefly illustrate here the construction which makes use of the quantum superalgebra ${\cal U}_q({\mathfrak {osp}}(1|2))$. Following \cite{kure} this quantum superalgebra is a deformation of the ${\mathfrak{osp}}(1|2)$ Lie superalgebra, recovered in the special limit $\eta\rightarrow 0$, where $\eta$ is a complex deformation parameter and $q$ can be expressed as $q=e^\eta$.\par
 ${\cal U}_q ({\mathfrak{osp}}(1|2))$ is generated by the three elements $H, F_\pm$ satisfying, in terms of the complex parameter $\eta\neq 0$, the (anti)commutation relations
\bea\label{quantumalgebra}
\relax [H, F_\pm ]_\eta&=& \pm \frac{1}{2} F_\pm,\nonumber\\
\{F_+, F_-\}_\eta &=& \frac{\sinh (\eta H)}{\sinh(2\eta)}=\frac{e^{\eta H}-e^{-\eta H}}{e^{2\eta}-e^{-2\eta}}.
\eea
In the limit when $\eta$ goes to zero one recovers the ordinary ${\mathfrak{osp}}(1|2)$ anticommutator $\{F_+,F_-\}$ of the $osp(1|2)$ odd generators:
\bea
\lim_{\eta\rightarrow 0}\{F_+,F_-\}_\eta &=&\{F_+,F_-\} =\frac{1}{2}H.
\eea
The quantum superalgebra ${\cal U}_q({\mathfrak{osp}}(1|2))$ has the structure of a graded Hopf superalgebra where, in particular, the following relations for the coproduct hold:
\bea
\Delta (H) &=& H\otimes {\bf 1}+{\bf 1}\otimes H,\nonumber\\
\Delta (F_\pm) &=& F_\pm \otimes e^{\frac{\eta}{2}H}+e^{-\frac{\eta}{2}H}\otimes F_\pm.
\eea
A single-particle Hilbert space ${\cal H}_\eta$ can be expressed as a lowest-weight representation of ${\cal U}_q({\mathfrak{osp}}(1|2))$, defined by the Fock vacuum $|0\rangle_\eta$ such that
\bea
H|0\rangle_\eta &=&\lambda |0\rangle_\eta,\nonumber\\
F_-|0\rangle_\eta &=&0,
\eea
where $\lambda$ is a given ``vacuum energy" and the Hilbert space ${\cal H}_\eta$ is spanned by the (possibly infinite) series of vectors $|n\rangle_\eta$:
\bea
|n\rangle_\eta &=& F_+^n|0\rangle_\eta, \qquad {\textrm{where}} \quad n=0,1,2,3, \ldots .
\eea 
A non-vanishing vector $|n\rangle_\eta$ is an eigenvector of $H$ with $\lambda+\frac{n}{2}$ eigenvalue:
\bea
H|n\rangle_\eta &=& (\lambda+\frac{n}{2})|n\rangle_\eta.
\eea
The connection with the energy spectrum of the single-particle Majorana qubit is performed in two steps. At first one sets $\lambda=0$ and $H_1:= 2H$ as the normalized single-particle Hamiltonian. Next, in order to recover from
 ${\cal U}_q({\mathfrak{osp}}(1|2))$ a $2$-dimensional Hilbert space, a suitable projector $P$ ($P^2=P$) should be applied. The two-dimensional finite Hilbert space ${\cal H}_\eta^{(1)}$ is obtained from ${\cal H}_\eta$ by applying the projector $P=diag(1,1,0,0,0\ldots)$, defined as
\bea\label{proj}
&P|0\rangle_\eta =|0\rangle_\eta, ~~~P|1\rangle _\eta = |1\rangle_\eta, ~~~~{\textrm{while}}~~  P|n\rangle_\eta =0 ~~~{\textrm{for}} ~~~ n\geq 2.&
\eea
It follows that ${\cal H}_\eta^{(1)}\subset {\cal H}_\eta$ {{is spanned by the $P$ eigenvectors with $+1$ eigenvalue, so that $|0\rangle_\eta, |1\rangle_\eta \in {\cal H}_\eta^{(1)}$.}}\\
In the multi-particle sectors the $(N+1)$-particle Hilbert space ${\cal H}_\eta^{(N+1)}$ is spanned by the vectors
\bea
&(P\otimes P\otimes \ldots \otimes P)\cdot {\widehat{\Delta^{(N)}(F_+^n)}} \cdot (|0\rangle_\eta\otimes|0\rangle_\eta\otimes\ldots \otimes |0\rangle_\eta),&
\eea
where the ${\widehat{\Delta^{(N)}}}$ coproduct acts on the tensor product of $N+1$ spaces.\par
This construction allows to reproduce, see \cite{topvoli} for details, the multiparticle spectrum of the braided Majorana qubits with the identification, for $t=-e^{2i\pi g}$ given in (\ref{gparametrization}),
\bea\label{tandeta}
t&=& e^{-\frac{\eta}{2}}, \qquad{\textrm{so that ~~~$\eta= -2\pi i (2g-1)$.}}
\eea

\section{Third scenario of braided Majorana qubits:\\ ~~~~~~ A new type of ``Volichenko metasymmetry".}

A third scenario for the multiparticle sector of braided Majorana qubits is realized when the braided tensor product $\otimes_{br}$, instead of simply (following \cite{maj}) {\it being declared} to be braided, is reconstructed from an ordinary $\otimes$ tensor product via the introduction of intertwining operators. \par
The basic example is given by the mappings 
\bea\label{matrixrepof}
 (\gamma~\otimes_{br}{\mathbb I}_2)\mapsto \gamma\otimes {\mathbb I}_2, &\qquad
 ({\mathbb I}_2~\otimes_{br}\gamma)\mapsto W_t\otimes \gamma,
\eea
which allow to recover the $\otimes_{br}$ braiding relation (\ref{braidingamma})
\bea
 ({\mathbb I}_2~\otimes_{br}\gamma)\cdot 
 (\gamma~\otimes_{br}{\mathbb I}_2)&\mapsto& (W_t\otimes\gamma)\cdot(\gamma\otimes {\mathbb I}_2)= (W_t\gamma)\otimes\gamma,\nonumber\\
 (\gamma~\otimes_{br}{\mathbb I}_2)\cdot ({\mathbb I}_2~\otimes_{br}\gamma)&\mapsto&
(\gamma\otimes {\mathbb I}_2)\cdot (W_t \otimes\gamma) = (\gamma W_t)\otimes\gamma,
\eea
provided that the $2\times 2$ 
intertwining operator $W_t$ satisfies the consistency condition
\bea\label{consistencyintertwining}
W_t\gamma &=& (-t) \gamma W_t.
\eea
A solution, expressed for the level-$s$ root of unity in terms of the  $t_s=-e^{\frac{2i\pi}{s}}$ position, is given by
\bea\label{solutionintertwining}
W_{t_s}&=& \cos(\frac{-\pi}{s} )\cdot {\mathbb I_2} + i \sin(\frac{-\pi}{s} )\cdot X,\qquad {\textrm{where $X=$ {\footnotesize  $\left(\begin{array}{cc} 1&0\\0&-1\end{array}\right)$.}}}
\eea
Indeed, we get
\bea
W_{t_s}\gamma &=& e^{\frac{2\pi i}{s} }\gamma W_{t_s}.
\eea
Following this prescription, the braided $2$-particle ($2P$) and $3$-particle ($3P$) creation operators
of the Majorana qubits can be respectively expressed (see \cite{topvoli} for details) as
\bea\label{braidedcr23}
2P &:& A_1^\dagger := \gamma\otimes {\mathbb I}_2, \quad \qquad A_2^\dagger :=W_{t_s} \otimes \gamma;  \nonumber\\
3P &:& B_1^\dagger := \gamma\otimes {\mathbb I}_2\otimes{\mathbb I}_2, \quad B_2^\dagger :=W_{t_s} \otimes \gamma\otimes{\mathbb I}_2,\quad
B_3^\dagger := W_{t_s}\otimes W_{t_s}\otimes \gamma.
\eea
For the $2$-particle sector, the braided fermionic creation building blocks $A_1^\dagger, A_2^\dagger$ and their respective conjugate matrices $A_1, A_2$  are $4\times 4$ matrices given by
{\footnotesize{\bea \label{nonstandard2p}
A_1^\dagger =\left(\begin{array}{cccc} 0&0&0&0\\0&0&0&0\\1&0&0&0\\0&1&0&0\end{array}\right),  &&
A_2^\dagger= \left(\begin{array}{cccc} 0&0&0&0\\e^{\frac{-i\pi}{s} }&0&0&0\\0&0&0&0\\0&0&e^{\frac{i\pi}{s} }&0\end{array}\right), \nonumber\\
A_1 =\left(\begin{array}{cccc} 0&0&1&0\\0&0&0&1\\0&0&0&0\\0&0&0&0\end{array}\right),  &&
A_2= \left(\begin{array}{cccc} 0&e^{\frac{i\pi}{s} }&0&0\\0&0&0&0\\0&0&0&e^{\frac{-i\pi}{s} }\\0&0&0&0\end{array}\right).
\eea}} 
Together with an even $4\times 4$ central charge $c$ defined as
\bea \label{centralcharge}
c&=& diag(1,1,1,1).
\eea
they close, as shown below, 
a generalized ``mixed-bracket" Heisenberg-Lie algebra.\par
~\par

This construction allows to make a connection with the notion of ``symmetries wider than supersymmetry" presented by Leites and Serganova in \cite{lese1} (see also \cite{lese2}); this notion concerns the existence of statistics-changing maps which do not preserve  the ${\mathbb Z}_2$-grading of ordinary Lie superalgebras.
Leites-Serganova introduced the concepts of ``metamanifolds" (as an extension of supermanifolds),
``metaspace" (as an extension of superspace) and ``metasymmetry" (as an extension of supersymmetry). 
As a concrete implementation of their proposal they investigated statistics-changing maps induced by nonhomogeneous subspaces of Lie superalgebras closed under the superbrackets. This leads to the notion of {\it Volichenko algebras} (after a theorem proved by Volichenko) which satisfy a condition known as metaabelianess; this means that, for any $x,y,z$ triple of operators, the identity
\bea\label{metaabel}
 [x,[y,z]]& =& 0,
\eea
 which involves ordinary commutators, is satisfied.  \par
Concerning Volichenko algebras, historical motivations and references leading to their introduction are found in \cite{{lese1},{lese2}}. A construction of Volichenko algebras as algebras of differential operators is found in
\cite{iye}, while a recent account with updated references is \cite{lei}.\par

An intriguing example of a class of Volichenko algebras was presented in \cite{lese2}. Denoted as ${\mathfrak{vgl}}_\mu(p|q)$, they are given by the ${\mathfrak{gl(p|q)}}$ space of $(p+q)\times(p+q)$ supermatrices. Under the ${\mathbb Z}_2$-decomposition ${\mathfrak{gl(p|q)}}={\mathfrak h}_0\oplus {\mathfrak h}_1$ into even/odd sectors, by fixing $\mu=a/b\in {\mathbb C}P^1$, a multiplication ${\mathfrak h}_1\times {\mathfrak h}_1\rightarrow {\mathfrak h}_0$ is introduced through the formula
\bea
(x,y)_\mu&=& a[x,y]+b\{x,y\}\qquad {\textrm{for any $x,y\in {\mathfrak{h}}_1$.}}
\eea
For $ab\neq 0$ the $(.,.)_\mu$ bracket is an interpolation of ordinary commutators and anticommutators.\par
~\par
This particular example was the starting point to explore the possibility of introducing  ``mixed brackets" to describe the parastatistics of the braided Majorana qubits. There were two main reasons for that:\\
{\it i}) a single-particle Majorana qubit is created/annihilated by the $\gamma,\beta$ generators in the odd sector of the ${\mathfrak{gl}}(1|1)$ superalgebra, as shown in formulas (\ref{4op}, \ref{anticomm}) and\\
{\it ii}) the building blocks of the $2$-particle creation/annihilation operators are accommodated in the odd sector
of a ${\mathfrak{gl}}(2|2)$ superalgebra. \par
These considerations led to the introduction in \cite{topvoli} of ``mixed-bracket" generalized Heisenberg-Lie algebras which close 
generalized Jacobi identities. We present here the simplest construction for the level-$s$ $2$-particle braided Majorana qubits obtained from the five $4\times 4$ matrices  presented in (\ref{nonstandard2p},\ref{centralcharge}).

\subsection{A generalized mixed-bracket Heisenberg-Lie algebra}

Let $X,Y$ be two operators. Their mixed-bracket, defined in terms of a $\vartheta_{XY}$ angle and denoted as $(X,Y)_{\vartheta_{XY}}$, is an interpolation of the ordinary $[X,Y]$ commutator and  $\{X,Y\}$ anticommutator. We can set
\bea\label{volimixedbracket}
(X,Y)_{\vartheta_{XY}} &:=& i \sin(\vartheta_{XY}) \cdot [X,Y] + \cos(\vartheta_{XY})\cdot \{X,Y\},
\eea
where $\vartheta_{XY}$ belongs, $mod~   2\pi$, to the interval $\vartheta_{XY}\in [-\pi,\pi[$.\\

For the special case of the  five $4\times 4$ matrices introduced in (\ref{nonstandard2p},\ref{centralcharge}), 
a consistent generalized mixed-bracket $2$-particle Heisenberg-Lie algebra is introduced as follows.\par
It is convenient at first to rename the generators as
\bea
&G_0:=c=diag(1,1,1,1),\quad G_{+1}:=A_1^\dagger,\quad G_{-1}:= A_1, \quad G_{+2} := A_2^\dagger, \quad G_{-2}:= A_2. &
\eea

The $5$ generators ($G_{\pm1}, G_{\pm 2}, G_0$) $2$-oscillator algebra has the only nonvanishing brackets given by
\bea
&(G_{\pm 1}, G_{\mp 1}) =  (G_{\pm 2}, G_{\mp 2}) = G_0.&
\eea
The mixed-bracket formulas, with the  explicit insertion of the $\vartheta_{IJ}$ angle dependence, are
\bea\label{anticommut}
&(G_{+1},G_{-1})_0=(G_{-1},G_{+1})_0= G_0, \qquad (G_{+1},G_{+1})_0=(G_{-1},G_{-1})_0 =0,&\nonumber\\
&(G_{+2},G_{-2})_0=(G_{-2},G_{+2})_0= G_0, \qquad (G_{+2},G_{+2})_0=(G_{-2},G_{-2})_0=0,&
\eea
together with 
\bea\label{genuinemixed}
&(G_{+1},G_{+2})_{+\frac{s+2}{2s}\pi}=(G_{+2},G_{+1})_{-\frac{s+2}{2s}\pi}= 0, &\nonumber\\
&(G_{+1},G_{-2})_{-\frac{s+2}{2s}\pi}=(G_{-2},G_{+1})_{+\frac{s+2}{2s}\pi}= 0, &\nonumber\\
&(G_{-1},G_{+2})_{-\frac{s+2}{2s}\pi}=(G_{+2},G_{-1})_{+\frac{s+2}{2s}\pi}= 0, &\nonumber\\
&(G_{-1},G_{-2})_{+\frac{s+2}{2s}\pi}=(G_{-2},G_{-1})_{-\frac{s+2}{2s}\pi}= 0. &
\eea

At $s=2$ these brackets define an ordinary, $2$ fermionic oscillators, Heisenberg-Lie algebra. \par
For any given $s=3,4,5,\ldots$, they are a mixed bracket generalization (a nontrivial interpolation of commutators/anticommutators) of the Heisenberg-Lie algebra which encodes the braid statistics of level $s$. \par
In the $s\rightarrow\infty $ untruncated limit one recovers again ordinary, i.e. not interpolated, commutators/anticommutators. In that limit,
on the other hand, one obtains the Heisenberg-Lie algebra of $2$ parafermionic oscillators, see \cite{topvoli} for details.

Since $G_0$ is a central element, the level$-s$ mixed-bracket algebras not only satisfy generalized Jacobi identities; they also satisfy a ``metaabelianess condition with respect to the mixed brackets". This means that
\bea\label{quantummetaabel}
(G_I,(G_J,G_K)) &=& 0 \qquad {\textrm{for any ~~$I,J,K=0,\pm 1,\pm 2$
}}.
\eea
The ordinary metaabelianess condition is not satisfied by $G_{\pm1}, G_{\pm 2}, G_0$. One can explicitly check, e.g., that
\bea
\relax [G_{+1},[G_{+2},G_{-2}]] &\neq &0.
\eea
One of the consequences is that braided Majorana qubits enlarge the notion of ``Volichenko algebras" as defined by Leites-Serganova. A key observation is that the Leites-Serganova construction is {\it classical} since  their notion of metaspace is applied to classical supergeometry and classical Lie superalgebras. On the other hand, the construction here discussed is related to quantum groups at roots of unity. This suggests the possibility of extending the notion of ``metaspace" to a ``quantum metaspace".\par

It is worth mentioning  that the construction of the ``mixed-bracket" Heisenberg-Lie algebras is performed for any 
$N$-particle sector and that the mixing angles $\vartheta_{IJ}$  which induce a closed algebraic structure are determined.

\section{Conclusions}

Section {\bf 2} presents the notion of {\it ${\mathbb Z}_2$-graded Majorana qubit}, while Sections {\bf 3}, {\bf 4} and {\bf 5} succintly summarize the results of \cite{{topqubits},{topvoli}} concerning braiding properties and mathematical structures describing them. The main features can be itemized as follows:\par
~\par
$\bullet$~~~The multiparticle quantization depends on a discrete control parameter $s=2,3,4,\ldots$ which labels a Gentile-type parastatistics where at most $s-1$  excited states,  as seen from  the (\ref{truncatedenergy}) energy spectra, are accommodated in an $N$-particle sector. \par
$\bullet$~~~The multiparticle quantization is determined by a graded Hopf algebra endowed with a braided tensor product.\par
$\bullet$~~~An alternative formulation, recovered from superselected reps of the quantum superalgebra ${\cal U}_q ({\mathfrak{osp}}(1|2))$, ``explains" the truncations of the (\ref{truncatedenergy}) energy spectra. \par
$\bullet$~~~Another alternative formulation is recovered from mixed-bracket generalizations of\\ Heisenberg-Lie algebras, where the mixed brackets are, see formulas (\ref{volimixedbracket}) and (\ref{genuinemixed}),  interpolations of ordinary commutators and anticommutators.\par
$\bullet$~~~The mixed-bracket structure is a quantum group generalization of the Leites-Serganova notion of {\it Volichenko algebras},  the classical notion of metaabelianess based, see (\ref{metaabel}), on ordinary commutators being now replaced by a metaabelianess with respect to the mixed brackets, see (\ref{quantummetaabel}).\par
~\par

I can now present a weaker form of the question asked in the Title. Let's reformulate it as:
\bea
&{\textrm{{\it Is there a minimal setting for a braided quantum model and, if so, which?}}}&
\eea
I claim that the first part of this question has a clear affirmative answer and that, for the second part, the
braided Majorana qubits  provide a minimal braided quantum model. The justification is presented here.

~\\
- It is well known that Leibniz, inspired by the hexagrams of the Chinese classic {\it I Ching},  introduced in the West
({\it Explication de l'arithm\'etique binaire}, published in 1703 under the spelling ``Leibnitz" in
 M\'emoires de l'Acad\'emie royale des Sciences, pages 83-89) the binary code represented by the digits $0$ and $1$. Leibniz advocated the advantages, due to its simplicity, of performing computations in this notation. 
Positional number systems  work with any base (number of digits) greater than one, like our $10$ notation, the Babylonian $60$, etc.. The base $2$ favoured by Leibniz is clearly the minimal choice since it involves the minimal number, $2$, of admissible digits. This simple minimal choice is the reason why ordinary computers manipulate bits and are engineered to work with Boolean logic.\par
~\\
- For ``ordinary" quantum computers there is not much to add to what already stated in the Introduction. A minimal quantum framework is based on the manipulation (and entanglement) of $2$-component states, the qubits.\par
~\\
- Concerning braided quantum mechanics satisfying a braided (anyonic) statistics, the Majorana qubits play the same role as ordinary qubits. It is sufficient to pinpoint similarities/differences of  ordinary qubits versus ${\mathbb Z}_2$-graded Majorana qubits. Let's have a closer look. \\
~\par
The ordinary qubit is described by the $2$-state vector 
{\small{\bea\label{qubit01ordinary}
{\textrm{qubit:}} &&~~~ \left(\begin{array}{c} a\\b\end{array}\right) , \quad {\textrm{with ~ $a,b\in{\mathbb C}$~  and ~$|a|^2+|b|^2>0$.}}
\eea}}
The inequivalent physical configurations, determined by the ray vector, are obtained from the normalization condition
$|a|^2+|b|^2=1$ plus a phase invariance. They are expressed by the Bloch sphere ${\bf S}^2$.\\
~\par

The ${\mathbb Z}_2$-graded Majorana qubit presents both bosonic (even) and fermionic (odd) states which, under a superselection rule, {\it cannot be linearly combined}. It is therefore described by the set of states
{\small{\bea\label{qubit01bis}
{\textrm{$\mathbb Z_2$-graded qubit:}} &&{\textrm{either even $~\left(\begin{array}{c} a\\0\end{array}\right)~$ or odd  $~\left(\begin{array}{c} 0\\b\end{array}\right)$,}}  ~~~{\textrm{with ~ $a,b\in{\mathbb C}$~  and ~$|a|^2+|b|^2>0$.}}
\eea}}
As recalled at the end of Section {\bf 2}, its inequivalent physical configurations are determined by the ray vectors and correspond to ${\bf Z}_2$, i.e. one bit of information.\\

${\mathbb Z}_2$-graded Majorana qubits share, with ordinary qubits, the same minimalistic properties;  they can  even be regarded as a simpler version which allows to introduce a non-trivial braiding for $s>2$. The (\ref{btmatrix}) braiding matrix $B_t$, parametrized by the inequivalent values $t\equiv t_s= e^{\frac{-2\pi i}{s}}$, satisfy idempotent relations in terms of powers of $s$:
\bea
B_{t_s}^s &=& {\mathbb I}_4.
\eea
The $s=2$ case corresponds to an ordinary bosons/fermions representation of the permutation group. In the mixed-bracket approach the nontrivial braiding is realized by nontrivial interpolations of commutators/anticommutators, see (\ref{genuinemixed}). Somewhat unexpectedly, in the $s\rightarrow\infty$ limit the interpolation of commutators/anticommutators disappears. In that limit one recovers \cite{topvoli} a Rittenberg-Wyler type of parafermions \cite{{rw1},{rw2}}. Unlike the anyonic braid statistics,  this class of paraparticles can live in any space dimension, see \cite{topisqs}. It has been shown in \cite{{top1},{top2}} that Rittenberg-Wyler paraparticles are theoretically detectable (see also the account in \cite{topijgmmp}).\\
The nontrivial braiding, which can in principle be applied to implement the Kitaev's proposal of topological quantum computation, is realized for any finite, integer value of $s$ in the range $s=3,4,5,\ldots$ . \\
In the multiparticle sectors the inequivalent physical configurations of braided Majorana qubits are determined, see \cite{topqubits}, by generalized Bloch spheres; for a $3$-component energy level graded Hilbert space the physical bosonic subspace is described by an ${\bf S}^2$ sphere and the fermionic subspace by a single point, while for a $4$-component energy level graded Hilbert space both bosonic and fermionic physical subspaces are described by an ${\bf S}^2$ sphere. Higher components imply more complicated structures.
\\
~\par
Braided Majorana qubits, albeit minimalistic, have a rich structure. They offer an ideal playground to test several mathematical frameworks applied in the construction of braided quantum models. Some of these frameworks, like the mixed-bracket Heisenberg-Lie algebras which generalize the notion of Volichenko algebras, have yet to be fully mathematically investigated. Different frameworks present different advantages. Indeed, one can note that the construction detailed in Section {\bf 3} involving a graded Hopf algebra endowed with a braided tensor product is more direct; the quantum group reps connection explained in Section {\bf 4}, on the other hand, clarifies the nature of the spectrum truncations observed in (\ref{truncatedenergy}).

Let's now come back to the (slightly rephrased) question asked in the Title:
\bea
&{\textrm{\it Are braided Majorana qubits a minimal setting for Topological Quantum Computation?~~~}}&
\eea
For the moment we cannot give a definite answer. Braided Majorana qubits are quite a novel quantum model and,
up to now, the investigations focused on their interrelated, underlying, mathematical structures. The proper analysis of the knot logic, see \cite{kau}, which could be encoded  within braided Majorana qubits has yet to be started.
This presentation is intended, in the light of the advocated minimalistic viewpoint, to advertise the relevance of forthcoming investigations in this topic.

Having identified a minimal model satisfying the consistency condition {\it i}) stated in the Introduction, the next two conditions {\it ii}) and {\it iii}) should be seriously addressed in forthcoming investigations.

~\\
{\large{\bf Acknowledgments}}\\
~\\
This work was supported by CNPq (PQ grant 308846/2021-4).

\end{document}